\pdfoutput=1

\documentclass[11pt,twoside,a4paper,cmspaper,final,collab]{cms-tdr}
\def\svnVersion{446756P}\def\cmsCernNoTag{CERN-EP-2017-158}\def\cmsCernDate{\today}\def\cmsMessage{Published in Physics Letters B as \href{http://dx.doi.org/10.1016/j.physletb.2017.12.069}{\doi{10.1016/j.physletb.2017.12.069.}}}

\begin{document}\cmsNoteHeader{SUS-16-045}

\hyphenation{had-ron-i-za-tion}
\hyphenation{cal-or-i-me-ter}
\hyphenation{de-vices}
\RCS$Revision: 446715 $
\RCS$HeadURL: svn+ssh://svn.cern.ch/reps/tdr2/papers/SUS-16-045/trunk/SUS-16-045.tex $
\RCS$Id: SUS-16-045.tex 446715 2018-02-19 06:02:53Z cpena $
\newlength\cmsFigWidth
\ifthenelse{\boolean{cms@external}}{\setlength\cmsFigWidth{\columnwidth}}{\setlength\cmsFigWidth{0.75\textwidth}}
\ifthenelse{\boolean{cms@external}}{\providecommand{\cmsLeft}{top\xspace}}{\providecommand{\cmsLeft}{left\xspace}}
\ifthenelse{\boolean{cms@external}}{\providecommand{\cmsRight}{bottom\xspace}}{\providecommand{\cmsRight}{right\xspace}}
\newlength\cmsTabSkip\setlength\cmsTabSkip{1.8ex}

\providecommand{\PSGcmpDo}{\ensuremath{\widetilde{\chi}^\mp_{1}}\xspace}

\newcommand{\MR}{\ensuremath{M_\mathrm{R}}\xspace}
\newcommand{\Rtwo}{\ensuremath{\mathrm{R}^2}\xspace}
\newcommand{\R}{\ensuremath{\mathrm{R}}\xspace}
\newcommand{\MRT}{\ensuremath{M_\mathrm{T}^\mathrm{R}}\xspace}

\cmsNoteHeader{SUS-16-045}
\title{Search for supersymmetry with Higgs boson to diphoton decays using the razor variables at $\sqrt{s}=13\TeV$}

\date{\today}

\abstract{An inclusive search for anomalous Higgs boson production in the diphoton decay channel and in association with at least one jet is presented, using LHC proton-proton collision data collected by the CMS experiment at a center-of-mass energy of 13\TeV and corresponding to an integrated luminosity of 35.9\fbinv. The razor variables $M_{\R}$ and ${\R}^2$, as well as the momentum and mass resolution of the diphoton system, are used to categorize events into different search regions. The search result is interpreted in the context of strong and electroweak production of supersymmetric particles. We exclude bottom squark pair-production with masses below 450\GeV for bottom squarks decaying to a bottom quark, a Higgs boson, and the lightest supersymmetric particle (LSP) for LSP masses below 250\GeV. For wino-like chargino-neutralino production, we exclude charginos with mass below 170\GeV for LSP masses below 25\GeV. In the GMSB scenario, we exclude charginos with mass below 205\GeV for neutralinos decaying to a Higgs boson and a goldstino LSP with 100\% branching fraction.
 }

\hypersetup{%
pdfauthor={CMS Collaboration},%
pdftitle={Search for supersymmetry with Higgs boson to diphoton decays using the razor variables at sqrt(s) = 13 TeV},%
pdfsubject={CMS},%
pdfkeywords={CMS, physics, Higgs}}

\maketitle
\section{Introduction}
\label{sec:intro}

\begin{figure*}
\centering
\includegraphics[width=0.4\textwidth]{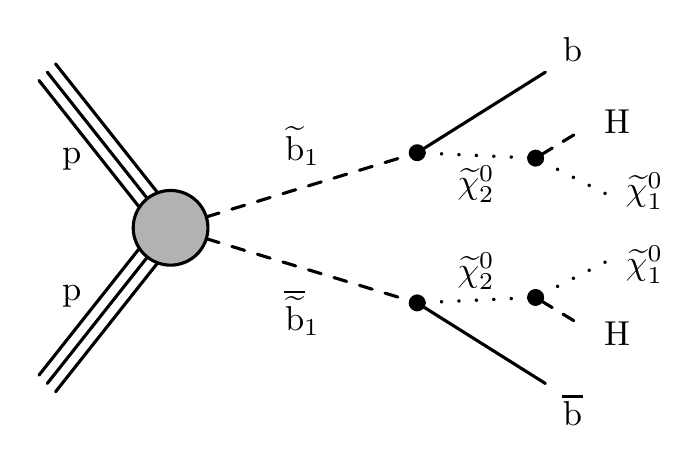}
\includegraphics[width=0.4\textwidth]{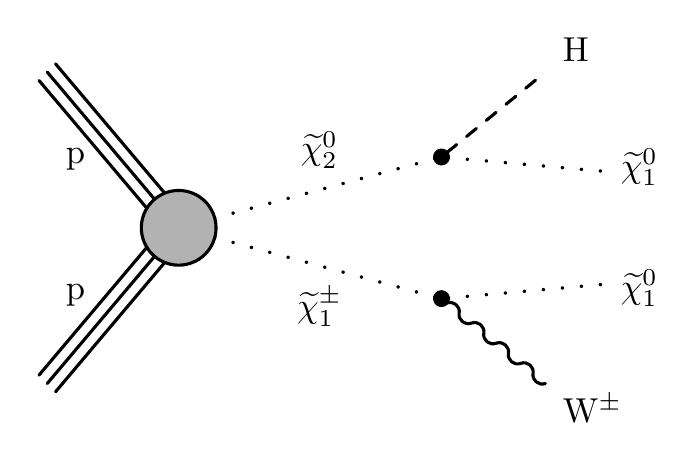}\\
\includegraphics[width=0.4\textwidth]{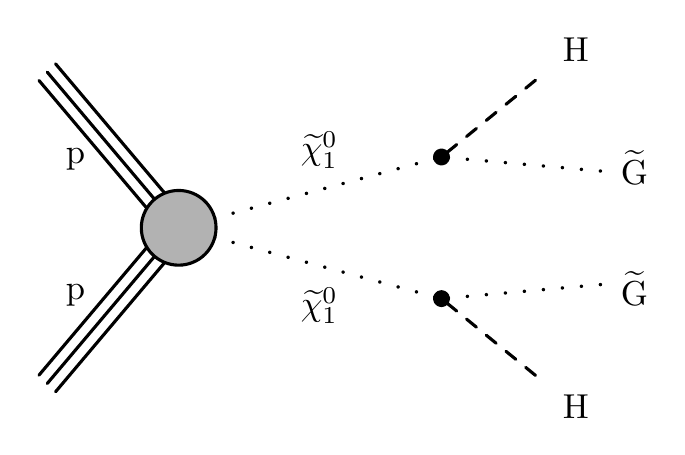}
\includegraphics[width=0.4\textwidth]{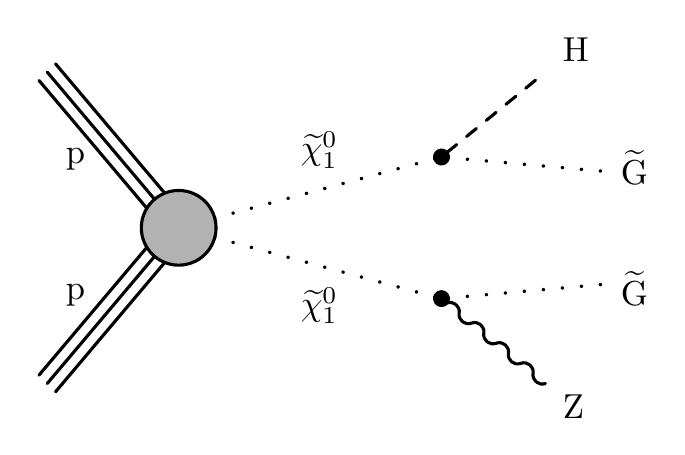}
\caption{Diagrams displaying the simplified models that are being considered. Upper left: bottom squark pair production; upper right: wino-like chargino-neutralino production; bottom: the two relevant decay modes for higgsino-like neutralino pair production in the GMSB scenario.
\label{fig:SMSDiagrams}}
\end{figure*}

The discovery of the Higgs boson~\cite{ATLASHiggs, CMSHiggs, CMSHiggslong}, the first fundamental scalar particle ever observed,
has opened a new window for exploring physics not described by the standard model (SM) of particle physics.
Many models of physics beyond the standard model (BSM) postulate the existence of cascade decays of heavy states involving Higgs
bosons~\cite{Monaco:2013poa,Duarte:2017bbq}. In the minimal supersymmetric standard model (MSSM)~\cite{Dimopoulos:1981zb}, Higgs
bosons may be produced in a variety of ways. The bottom squark, the supersymmetric partner of the bottom quark, produced via the
strong interaction, may decay to a Higgs boson, quarks, and the lightest supersymmetric particle (LSP). Similarly charginos or
neutralinos produced through the electroweak interaction may decay to a Higgs boson and the LSP. Of particular interest are
scenarios with gauge-mediated supersymmetry breaking (GMSB), where the lightest neutralino may decay to a Higgs boson and the
goldstino LSP~($\PXXSG$)~\cite{Dimopoulos:1996vz,Matchev:1999ft}. The decay signature depends on whether the chargino and neutralino
mixed states are dominated by the wino or higgsino components, the respective supersymmetric partners of the W and Higgs bosons.
Diagrams of simplified models~\cite{Alves:2011wf} for the scenarios considered are shown in Fig.~\ref{fig:SMSDiagrams}.
In this paper, we denote the Higgs boson as H to indicate that it is the particle observed by the ATLAS and CMS experiments.
In the MSSM, this particle is typically assumed to correspond to the lighter of the two CP-even Higgs particles and is often
denoted as h. For the GMSB scenario, we consider simplified models where Higgsino-like charginos and neutralinos are nearly mass-degenerate
and both chargino-neutralino and neutralino-pair production result in very similar final state signatures, and are hereafter
collectively referred to as chargino-neutralino production in this paper.
These examples of BSM production of Higgs bosons motivate an inclusive search for anomalous Higgs boson production that is
broadly sensitive to a wide range of supersymmetric (SUSY) scenarios. Similar searches for supersymmetric particles decaying to
Higgs bosons have been performed by the ATLAS and CMS collaborations in the past using 8\TeV collision data
and can be found in references~\cite{Aad:2015jqa,Khachatryan:2014mma,CMS-PAS-SUS-14-017}.

In this Letter, we present an updated search for supersymmetry events with at least one Higgs boson candidate decaying to two photons
produced in association with at least one jet produced in 13\TeV proton-proton collisions. The data were collected by the CMS
experiment and correspond to an integrated luminosity of 35.9\fbinv~\cite{CMS-PAS-LUM-17-001}. The diphoton decay mode of the Higgs boson provides a good compromise between branching fraction and background rejection. The transverse momentum of the Higgs boson candidate, the expected mass resolution, and the razor variables \MR and \Rtwo~\cite{razorPRL, razorPRD}, explained in detail in Section~\ref{sec:selection}, are used to define event categories which generically enhance BSM signals relative to SM backgrounds. The potential signal is extracted from the dominant nonresonant multijet background through a fit to the diphoton mass distribution. The results of the search are interpreted in terms of simplified models of bottom squark pair production and chargino-neutralino production.

\section{The CMS detector, trigger, and event reconstruction}\label{cmsdetector}

The central feature of the CMS detector is a superconducting solenoid of 6\unit{m} internal diameter, providing a magnetic field of 3.8\unit{T}. Within the solenoid volume are a silicon pixel and strip tracker, a lead tungstate crystal electromagnetic calorimeter (ECAL), and a brass and scintillator hadron calorimeter (HCAL), each composed of a barrel and two endcap sections. Extensive forward calorimetry complements the coverage
provided by the barrel and endcap detectors. Muons are measured in gas-ionization detectors
embedded in the magnet steel flux-return yoke outside the
solenoid. A more detailed description of the CMS detector, together with a definition
of the coordinate system used and the relevant kinematic variables, can be found in Ref.~\cite{Adolphi:2008zzk}.

Signal event candidates are recorded using a diphoton trigger, requiring the transverse energy of the leading and subleading photons to be larger than 30\GeV and 18\GeV, respectively and their invariant mass to be larger than 90\GeV. Additional requirements on the photon shower shape and isolation are imposed to reduce the background rate and improve the signal purity~\cite{CMS-PAS-HIG-16-040}. The efficiency of the trigger with respect to events passing the offline selection is measured to be above 98\%.

Physics object candidates are reconstructed using a global event description based on the CMS particle-flow (PF)
algorithm~\cite{CMS-PRF-14-001}, which
identifies particles through an optimized combination of information from the various detector subsystems.
Photon candidates are selected by imposing ``loose'' requirements on the shower shape in the electromagnetic
calorimeter, the ratio of energy measured in the HCAL to the energy measured in the ECAL,
and the isolation in a cone around the direction of the
photon momentum~\cite{Khachatryan:2015iwa}. The isolation requirement is corrected for the effect of multiple proton
collisions in the same or adjacent bunch crossing (pileup) by subtracting the average energy
from pileup deposited in the isolation cone, estimated by averaging
the energy density over the event. Furthermore, photon candidates are rejected if they match an electron candidate that is not consistent with one leg of a conversion.
The photon selection efficiency was measured to be about 90\%~\cite{CMS-DP-2017-004} using tag and probe methods. The measured energy of photons is corrected for clustering and local
geometric effects using an energy regression trained on
 Monte Carlo (MC) simulation~\cite{Khachatryan:2015iwa}. This regression gives a significant improvement in the energy resolution of the photons (about 30\%) and provides an estimate of the uncertainty of the energy measurement that is used to separate events into high- and low-resolution categories.

The reconstructed vertex with the largest value of summed physics-object $\pt^2$ is taken to be
the primary $\Pp\Pp$ interaction vertex (PV). The physics objects used in this context are the objects
returned by a jet finding algorithm~\cite{Cacciari:2008gp,Cacciari:2011ma} applied to all charged
tracks associated with the vertex under consideration, plus the corresponding associated missing transverse
momentum.

The charged PF candidates associated with the PV and the neutral PF candidates are clustered
into jets using the anti-\kt algorithm~\cite{Cacciari:2008gp} with a distance parameter $R=0.4$,
as implemented in the \FASTJET package~\cite{Cacciari:2011ma}. The jet momentum is determined as
the vectorial sum of all particle momenta in the jet. Jet energy corrections are derived based
on a combination of simulation studies, accounting for the nonlinear detector response and
the presence of pileup, together with in-situ measurements of the energy balance in dijet and
$\gamma$+jet events using the methods described in Ref.~\cite{Khachatryan:2016kdb}.
For this analysis, jets with $\abs{\eta}<3.0$ that
do not overlap with any identified photon are selected by
requiring $\Delta R = \sqrt{\smash[b]{(\Delta \eta)^2 + (\Delta \phi)^2}}>0.5$ between photon and jet candidates.
The combined secondary vertex (CSVv2) tagging algorithm~\cite{CMS-PAS-BTV-15-001} is used to identify
jets originating from the hadronization of b quarks. A loose working point is chosen that yields an
efficiency above 80\% and a mistag rate for light-flavor jets that is approximately 10\%. The negative
vector sum of the reconstructed $\pt$ of all PF candidates in an event defines the missing transverse momentum
$\ptvecmiss$ in the event, and its magnitude is referred to as $\pt^{\text{miss}}$.
Events with detector- and beam-related noise that can mimic event topologies
with high energy and large $\pt^{\text{miss}}$ are filtered out by use of dedicated noise
reduction algorithms~\cite{Chatrchyan:2011tn,Chatrchyan:2012lia,Khachatryan:2014gga}.

\section{Event simulation}\label{sec:data}

Simulated event samples are used to model the SM Higgs backgrounds
in the search regions, and to calculate the selection efficiencies for SUSY signal models.
Samples of SM Higgs production via gluon fusion, vector boson fusion,
associated production with a $\PW$ or a $\cPZ$ boson, $\bbbar\PH$, and $\ttbar\PH$ are generated
using the next-to-leading order (NLO) \MGvATNLO~v2.2.2~\cite{Alwall:2014hca} event generator.
The Higgs mass is assumed to be 125.0\GeV for the simulated event samples and is within the
uncertainty of the currently best measured value~\cite{Aad:2015zhl}.
For the gluon fusion production mode, the sample is generated with
up to two extra partons to model initial-state radiation (ISR) calculated at the matrix element
level with NLO accuracy with the matching scheme described in reference~\cite{Frederix:2012ps}.
The SUSY signal MC samples are generated using \MGvATNLO at leading order accuracy
with up to two extra partons in the matrix element calculations with the matching scheme
described in reference~\cite{Alwall:2007fs}. In both cases, \PYTHIA v8.2~\cite{Sjostrand2008852}
is used to model the fragmentation and parton showering with the CUETP8M1 tune~\cite{Skands2014}.
The {NNPDF3.0LO} and {NNPDF3.0NLO}~\cite{Ball:2014uwa} parton distribution
functions (PDFs) are used for the LO and NLO accuracy generators respectively. The SM Higgs background
and bottom squark pair-production signal samples are simulated using a \GEANTfour-based model~\cite{geant4}
of the CMS detector, while the chargino-neutralino and neutralino-pair production signal samples
are simulated with the CMS fast simulation package~\cite{FastSim}.
While generally providing an accurate description, the fast simulation does sometimes yield inaccurate
predictions of the missing transverse momentum tail. These inaccuracies are accounted for by larger systematic
uncertainties for the signal yield prediction in the relevant phase space estimated as the difference
between signal yields predicted using the generator level missing transverse momentum and the
missing transverse momentum reconstructed based on the fast simulation.
All simulated events include the effects of pileup and are processed with the same chain of
reconstruction programs used for collision data.

To improve the \MADGRAPH modeling of ISR in the SUSY signal MC samples, we apply a correction as a function of the
multiplicity of ISR jets for sbottom pair production, and as a function of the transverse momentum ($\pt^{\mathrm{ISR}}$)
of the chargino-neutralino system for chargino-neutralino production, derived from studies of \ttbar and Z+jets
events, respectively. The correction factors vary between 0.92 and 0.51 for ISR jet multiplicity between one
and six, and between 1.18 and 0.78 for $\pt^{\mathrm{ISR}}$ between 125 and 600\GeV. The correction has a
small effect on the signal yields at the level of about $1\%$. The full size of
this correction is taken as a systematic uncertainty.

The Higgs production cross sections are obtained from the recommendations of the LHC
Yellow Report 4 of the LHC Higgs Cross Section Working Group~\cite{deFlorian:2016spz}.
The SUSY signal production cross sections are calculated to NLO plus next-to-leading logarithmic (NLL)
accuracy~\cite{NLONLL1,NLONLL2,NLONLL3,NLONLL4,NLONLL5,Borschensky:2014cia},
assuming all SUSY particles other than those in the relevant diagram to be too heavy to participate in the interaction. These NLO+NLL cross sections and their associated uncertainties~\cite{Borschensky:2014cia} are used to derive the exclusion limits on the masses of SUSY particles.

\section{Event selection and search categories}
\label{sec:selection}

We select events with two photons that satisfy the selection criteria described above. Both photons must be in the barrel region of the electromagnetic calorimeter, with $\abs{\eta} < 1.44$, and have $\pt>20$\GeV. At least one of the two photons must have $\pt>40$\GeV. If multiple photon pairs are identified, the pair with the largest scalar sum of the transverse momenta is chosen as the Higgs boson candidate in the event. The Higgs boson candidate
mass is required to be between 103\GeV and 160\GeV in order to cover a sufficiently large background dominated sideband region.

The Higgs boson candidate and any additional identified jets with $\pt>30$\GeV and $\abs{\eta}<3.0$ are clustered into two
hemispheres (megajets) according to the Razor megajet algorithm~\cite{razorPRD},
which minimizes the sum of the squared-invariant-mass values of the two megajets.
To converge, the algorithm requires at least one such identified jet in the event.
Next, the razor variables~\cite{razorPRL} \MR and \Rtwo are computed as follows:
\begin{align}
 \label{eq:MRstar}
 \MR &\equiv \sqrt{(\abs{\vec{p}^{j_{1}}}+\abs{\vec{p}^{j_{2}}})^2 -({p}^{j_{1}}_z+{p}^{j_{2}}_z)^2},\\
\Rtwo &\equiv \left( \frac{\MRT}{\MR} \right)^2,\label{eq:Rtwo}
\end{align}
where $\vec{p}$ is the momentum of a megajet, $p_z$ is its longitudinal component,
and $j_{1}$ and $j_{2}$ are used to label the two megajets. In the definition of \Rtwo,
the variable $\MRT$ is defined as:
\begin{equation}
\MRT \equiv \sqrt{ \frac{\pt^{\text{miss}}(\pt^{j_{1}}+\pt^{j_{2}}) -\ptvecmiss \cdot (\ptvec^{\,j_{1}}+\ptvec^{\,j_{2}}) }{2}}.\label{eq:MRT}
\end{equation}
The razor variables \MR and \Rtwo provide discrimination between SUSY signal models and SM background processes with SUSY signals typically having large values of \MR and \Rtwo, while the SM background exhibits an exponentially falling
spectrum in both variables.

\begin{figure}
\centering
\includegraphics[width=0.5\textwidth]{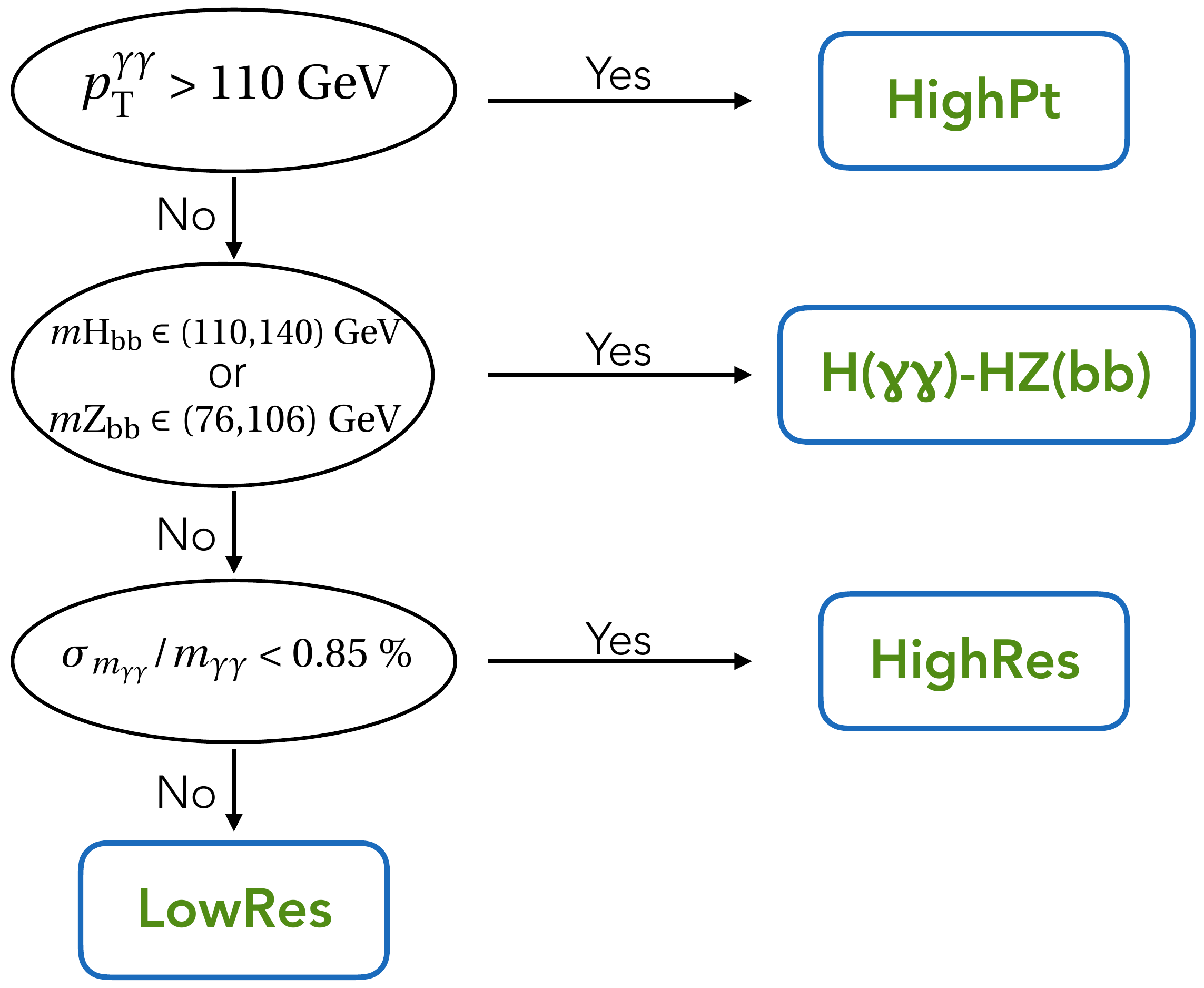}
\caption{A flowchart showing the event categorization procedure.\label{fig:CategorizationFlowchart}}
\end{figure}

The selected events are separated into four mutually exclusive categories. An event is categorized as ``HighPt'' if the transverse momentum
of the selected Higgs boson candidate is larger than 110\GeV. Otherwise, if the event contains two b-tagged jets whose invariant
mass is between 76 and 106\GeV, or between 110 and 140\GeV, it is categorized as ``$\PH(\gamma\gamma)$-$\PH\Z(\PQb\PQb)$''.
The remaining events are categorized as ``HighRes'' and ``LowRes'' if the diphoton mass resolution estimate $\sigma_{M}/M$ is
smaller or larger than 0.85\%, respectively, where $\sigma_{M}$ is computed
as:
\begin{equation}
\sigma_{M} = \frac{1}{2}\,\sqrt{(\sigma_{E\gamma 1}/E_{\gamma 1})^{2} + (\sigma_{{E}\gamma 2}/E_{\gamma 2})^{2}},
\end{equation}
where $E_{\gamma 1,2}$ is the energy of each photon and $\sigma_{{E}\gamma 1,2}$ is the estimated energy resolution for each photon.
A graphical representation of the categorization procedure is shown in Fig.~\ref{fig:CategorizationFlowchart}.
The ``HighPt'' category isolates SUSY events producing high-\pt Higgs bosons; the ``$\PH(\gamma\gamma)$-$\PH\Z(\PQb\PQb)$'' category
isolates SUSY signals that produce two Higgs bosons or a Higgs boson and a Z boson in the final state; and the HighRes and
LowRes categories further improve the discrimination between signal and background in the remaining event sample.
The ``$\PH(\gamma\gamma)$-$\PH\Z(\PQb\PQb)$'' category combines events with two Higgs bosons or a Higgs boson and a Z boson
in order to achieve a sufficiently large number of events in the sideband for the background estimation method described in
Section~\ref{sec:bkg} to remain unbiased.

Each event category is further divided into bins in the \MR and \Rtwo variables, which define the exclusive
search regions. A significant excess of events above the SM expectation in one or several bins would
provide evidence of BSM physics. The search regions are chosen based on an
optimization procedure that maximizes the expected sensitivity
to the simplified bottom squark pair production model discussed further in Section~\ref{sec:results},
and are summarized in Table~\ref{tab:binSummary}.
The bins in the \MR and \Rtwo variables are kept identical for the HighRes and LowRes categories to
allow for simultaneous signal extraction, since the ratio of the event yields in these two categories does not
depend on the details of the signal model.

\begin{table*}[htb]
\centering
\topcaption{A summary of the search region bins in each category is presented. The functional form used to model the nonresonant background is also listed. An exponential function of the form $\re^{-a m_{\gamma\gamma}}$ is denoted as ``single-exp''; a linear combination of two independent exponential functions of the form $\re^{-a m_{\gamma\gamma}}$ and $\re^{-b m_{\gamma\gamma}}$ is denoted as ``two-exp''; a modified exponential function of the form $\re^{-a m_{\gamma\gamma}^{b}}$ is denoted as ``mod-exp''; and a Bernstein polynomial of degree $n$~\cite{Khachatryan:2014ira} is denoted by ``poly-n''. The bin labels 9--13 are used for both the HighRes and LowRes categories because the data in these categories are always fitted simultaneously with potentially different nonresonant background models used. Further details on the simultaneous fit are discussed in Section~\ref{sec:bkg}. }
\begin{tabular}{ccccl}
\hline
Bin number & Category & \MR (\GeVns{}) & \Rtwo   & Nonresonant bkg. model   \\
\hline
0          & HighPt   &  $\ge$600     &   $\ge$0.025          & single-exp   \\
1          & HighPt &  150--600    &   $\ge$0.130          & single-exp     \\
2          & HighPt &  $\ge$1250    &   0.000--0.025        & single-exp     \\
3          & HighPt &  150--450    &   0.000--0.130        & poly-3         \\
4          & HighPt &  450--600    &   0.000--0.035        & poly-3         \\
5          & HighPt &  450--600    &   0.035--0.130        & single-exp     \\
6          & HighPt &  600--1250   &   0.000--0.015        & two-exp        \\
7          & HighPt &  600--1250   &   0.015--0.025        & single-exp\\[\cmsTabSkip]
8          & $\PH(\gamma\gamma)$-$\PH\Z(\PQb\PQb)$   &  $\ge$150    &   $\ge$0.0    & single-exp\\[\cmsTabSkip]
\multirow{2}{*}{9}  & HighRes & 150--250     &   0.000--0.175        & mod-exp \\
                    & LowRes & 150--250      &   0.000--0.175         & poly-3        \\
\multirow{2}{*}{10} & HighRes & 150--250     &   $\ge$0.175          & single-exp     \\
                    & LowRes & 150--250     &   $\ge$0.175           & single-exp     \\
\multirow{2}{*}{11} & HighRes & $\ge$250     &   $\ge$0.05           & single-exp     \\
                    & LowRes & $\ge$250     &   $\ge$0.05            & poly-2         \\
\multirow{2}{*}{12} & HighRes & 250--600     &   0.000--0.05         & poly-2         \\
                    & LowRes & 250--600     &   0.000--0.05          & mod-exp        \\
\multirow{2}{*}{13} & HighRes & $\ge$600     &   0.000--0.05         & single-exp     \\
                    & LowRes & $\ge$600     &   0.000--0.05          & single-exp   \\
\hline
\end{tabular}
\label{tab:binSummary}
\end{table*}

\section{Background prediction}\label{sec:bkg}

There are two main classes of background events that pass the search selection criteria: SM Higgs production and nonresonant photon production, with either two promptly produced photons
or one prompt photon and one jet that is wrongly identified as a photon. The SM Higgs background is estimated from the MC simulation, while the nonresonant background prediction is estimated using a fit to the diphoton mass distribution observed in data.

Within each search bin, we extract a potential signal by performing an unbinned extended maximum likelihood fit to the diphoton mass spectrum. An example of such a fit is shown in Fig.~\ref{fig:ExampleFit}. The nonresonant background is modeled with a decaying functional form given in Table~\ref{tab:binSummary} for each individual search region bin. All parameters of the function are unconstrained in the fit. The functional form of the model used for each search region bin is selected on the basis of its Akaike information criterion (AIC) score~\cite{AIC}, which quantifies the trade-off between goodness-of-fit and model complexity. Each functional form is tested for fit biases with respect to a set of alternative models, all of which adequately describe the data in the diphoton mass sideband region (103--121\GeV and 129--160\GeV). The shape of the Higgs boson resonance from SM Higgs production and from decays of SUSY signals is modeled with a double Crystal Ball function~\cite{Oreglia:1980cs,Gaiser:1982yw} with two independent tail parameters that is fitted to the diphoton mass distribution obtained from the MC simulation. The parameters of each double Crystal Ball function are held constant
in the signal extraction fit procedure, with the exception of the parameter controlling the location of the peak, which is discussed further in Section~\ref{sec:systematics} below. The normalization of the SM Higgs boson background in each bin is predicted from the MC simulation and is constrained to that value in the fit within uncertainties. For the HighRes and LowRes categories, bins in the \MR and \Rtwo variables are fitted simultaneously. For a given search bin, the relative yields in the HighRes and LowRes categories are observed in the simulation to be largely process independent and are therefore constrained according to the simulation prediction. Based on these independent fits in each search bin, we obtain a model-independent search result, which can be used to evaluate whether the yield in any bin exhibits a statistically significant deviation from the background prediction.

We also perform a combined simultaneous fit using all of the search bins, to test specific SUSY simplified model signal hypotheses. In the combined fit, the yield in each bin for the SM Higgs background and the signal model under test are constrained to the MC simulation predictions within uncertainties. These uncertainties are modeled by use of nuisance parameters that account for various theoretical and instrumental uncertainties that can affect the SM Higgs boson background and SUSY
signal normalization, and are profiled in the fit. A more detailed discussion of systematic uncertainties can be found below in Section~\ref{sec:systematics}.
The MC simulation predictions for the SM Higgs boson background
normalization are shown in Table~\ref{tab:SMHBkgPrediction} for each bin in the search region.

\begin{table*}[htb]
\centering
\topcaption{The predicted yields for an example SUSY signal and the SM Higgs boson background processes for each search region are shown for an integrated luminosity corresponding to 35.9\fbinv. The signal yields given assume a bottom squark mass of 500\GeV and an LSP mass of 1\GeV. The contributions from each SM Higgs boson process are shown separately, and the total is shown in the rightmost column, along with its full uncertainty. The bin labels 9--13 are used for both the HighRes and LowRes categories as they are always fitted simultaneously.
\label{tab:SMHBkgPrediction}
}
\newcolumntype{x}{D{,}{\,\pm\,}{2.2}}
\
\begin{tabular}{ccccccccx}
\hline
 \multicolumn{2}{c}{}       & \multicolumn{1}{c}{Signal Yield}                           & \multicolumn{6}{c}{Expected SM Higgs yield} \\ \cline{4-9}\\[-1.8ex]
\multicolumn{1}{c}{Bin} & \multicolumn{1}{c}{Category} & \multicolumn{1}{c}{$\PSQb\PASQb$,$\PSQb \to \cPqb\PH\PSGczDo$} & \multicolumn{1}{c}{ggH} & \multicolumn{1}{c}{$\ttbar\PH$} & \multicolumn{1}{c}{VBF H} & \multicolumn{1}{c}{VH} & \multicolumn{1}{c}{$\bbbar\PH$} & \multicolumn{1}{c}{Total}\\
\hline
0 & HighPt & 10.2 & 3.4 & 1.4 & 0.49 & 0.78 & 0.02 & 6.1,1.2 \\
1 & HighPt & 0.2 & 1.7 & 0.58 & 0.18 & 1.8 & 0.01 & 4.3,0.8 \\
2 & HighPt & 5.7 & 5.1 & 0.64 & 2.5 & 0.17 & 0.04 & 8.5,1.7 \\
3 & HighPt & 0.2 & 55 & 0.96 & 11 & 6.3 & 0.41 & 74,21 \\
4 & HighPt & 0.6 & 17 & 0.50 & 4.7 & 1.1 & 0.18 & 23,7 \\
5 & HighPt & 0.6 & 3.5 & 0.61 & 0.55 & 0.59 & 0.04 & 5.3,1.2 \\
6 & HighPt & 5.7 & 19 & 0.80 & 7.6 & 0.82 & 0.15 & 29,8 \\
7 & HighPt & 4.0 & 5.4 & 0.46 & 1.1 & 0.45 & 0.02 & 7.4,2.1\\[\cmsTabSkip]
8  & $\PH(\gamma\gamma)$-$\PH\Z(\PQb\PQb)$ & 0.9 & 0.76 & 1.3 & 0.12 & 0.25 & 0.19 & 2.6,0.4\\[\cmsTabSkip]
\multirow{2}{*}{9}  & HighRes & 0.0 & 60 & 0.24 & 7.6 & 4.4 & 0.89 & 75,22 \\
                    & LowRes & 0.0 & 30 & 0.11 & 3.8 & 2.3 & 0.50 & 36,11 \\
\multirow{2}{*}{10} & HighRes & 0.0 & 1.1 & 0.12 & 0.12 & 0.46 & 0.02 & 1.8,0.6 \\
                    & LowRes & 0.0 & 0.44 & 0.07 & 0.08 & 0.27 & 0.01 & 0.9,0.2 \\
\multirow{2}{*}{11} & HighRes & 0.3 & 3.0 & 0.73 & 0.54 & 0.55 & 0.13 & 5.0,1.4 \\
                    & LowRes & 0.1 & 1.8 & 0.38 & 0.29 & 0.22 & 0.06 & 2.7,0.8 \\
\multirow{2}{*}{12} & HighRes & 0.1 & 37 & 0.66 & 8.9 & 1.4 & 0.83 & 50,14 \\
                    & LowRes & 0.1 & 21 & 0.33 & 4.7 & 0.79 & 0.42 & 26,6 \\
\multirow{2}{*}{13} & HighRes & 1.0 & 5.0 & 0.50 & 3.1 & 0.21 & 0.21 & 9.1,2.7 \\
                    & LowRes & 0.5 & 3.3 & 0.29 & 1.5 & 0.13 & 0.10 & 5.2,1.5\\
\hline
\end{tabular}
\end{table*}

\section{Systematic uncertainties}
\label{sec:systematics}

There are broadly two types of systematic uncertainties.
The first and dominant systematic uncertainty is in the shape and normalization of the nonresonant background. This is
propagated by profiling the normalization and shape parameters of the nonresonant background
functional form in an unconstrained way. The second and subdominant type of systematic uncertainty is in the
predictions of the SM Higgs background in the various search bins. These shape uncertainties are propagated
through the use of several independent nuisance parameters, where both theoretical and
instrumental effects are taken into account. The nuisance parameters are
constrained in the fit using log-normal prior functions, whose width reflects
the size of the systematic uncertainty. The independent systematic effects considered include missing
higher-order corrections, PDFs, trigger and selection efficiencies, jet energy
scale uncertainties, b tagging efficiencies, and the uncertainty in the integrated luminosity.
The uncertainty due to jet energy resolution uncertainties were also estimated and were found to be
negligible. The typical size of these effects on the expected limit is summarized in Table~\ref{tab:SignalSystematics}. Due to
effects of pileup and transparency loss in the ECAL crystals, we observe some simulation mismodeling of
the estimated mass resolution, which results in a systematic uncertainty of 10--24\% in the prediction
of the SM Higgs background and SUSY signal yields in the HighRes and LowRes event categories.
The systematic uncertainty in the photon energy scale is implemented as a Gaussian-distributed nuisance
parameter that shifts the Higgs boson mass peak position, constrained in the fit to lie within approximately 1\%
of the nominal Higgs boson mass observed in simulation. The systematic uncertainty for the modeling of the ISR
for the signal process is also propagated and is below $1\%$. For chargino-neutralino and neutralino-pair
production signal processes, the fast simulation was used to predict signal yields and an additional
systematic uncertainty is propagated for inaccuracies in the modeling of the missing transverse
momentum tail. This systematic uncertainty ranges between $1\%$ and $34\%$ depending on the search region bin.

\begin{table}
\centering
\topcaption{Summary of systematic uncertainties on the SM Higgs background and signal yield predictions,
and the size of their effect on the signal yield.}
\label{tab:SignalSystematics}
\begin{tabular}{lc}
\hline
Uncertainty source & Size (\%) \\
\hline
Integrated Luminosity                       & 2.5 \\
PDFs/renormalization/factorization scales   & 15--30\\
Trigger and selection efficiency                    & 3 \\
Jet energy scale                                    & 1--5 \\
Photon energy scale                                 & 1 \\
$\sigma_{M}/M$ categorization                       & 10--24 \\
b tagging efficiency                                & 4 \\
ISR modeling (signal only)                          & 1\\
Fastsim $\pt^{\text{miss}}$ modeling (signal only)    & 1--34\\
\hline
\end{tabular}
\end{table}

\section{Results and interpretations}
\label{sec:results}

\begin{figure}[htb]
\centering
\includegraphics[width=0.5\textwidth]{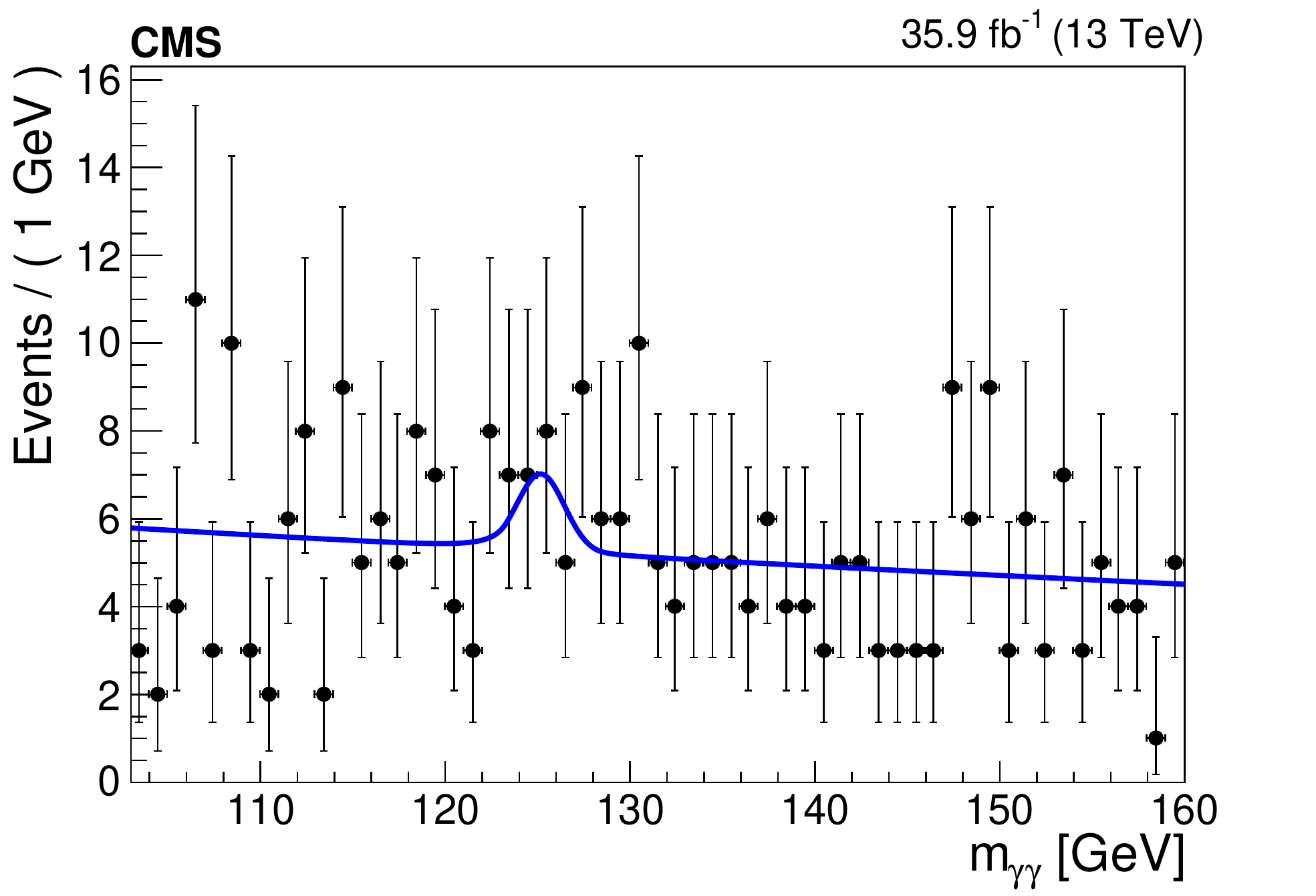}
\includegraphics[width=0.45\textwidth]{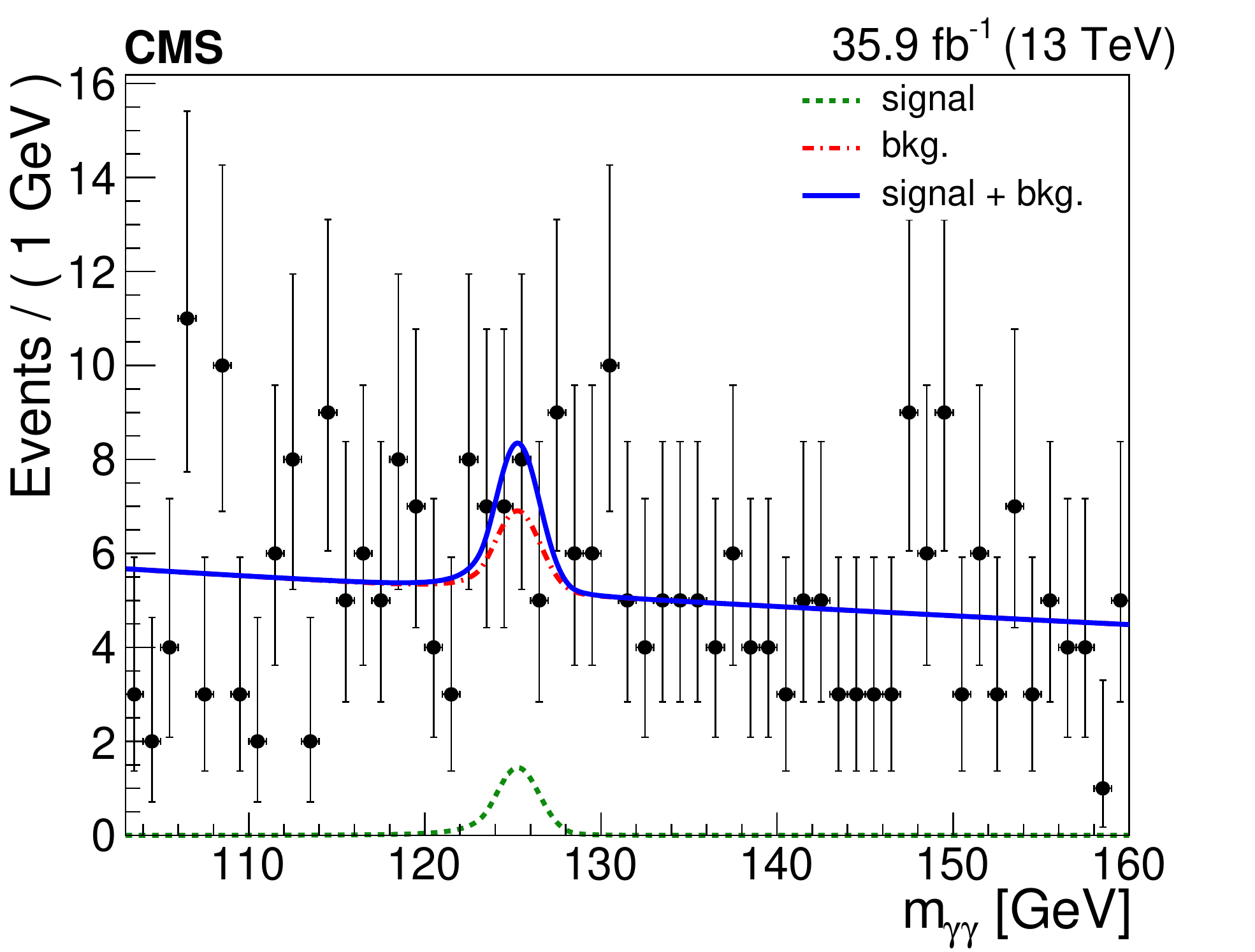}
\caption{The diphoton mass distribution in the search region bin with $\MR > 600$\GeV and $\Rtwo > 0.025$
in the HighPt category, along with the background-only fit (\cmsLeft) and the signal-plus-background fit (\cmsRight).
The red dot-dashed curve represents the fitted background prediction; the green dashed curve represents the best-fit signal;
and the blue solid curve represents the sum of the best-fit signal and the background.
\label{fig:ExampleFit}}
\end{figure}

\begin{figure}[h!]
\centering
\includegraphics[width=\cmsFigWidth]{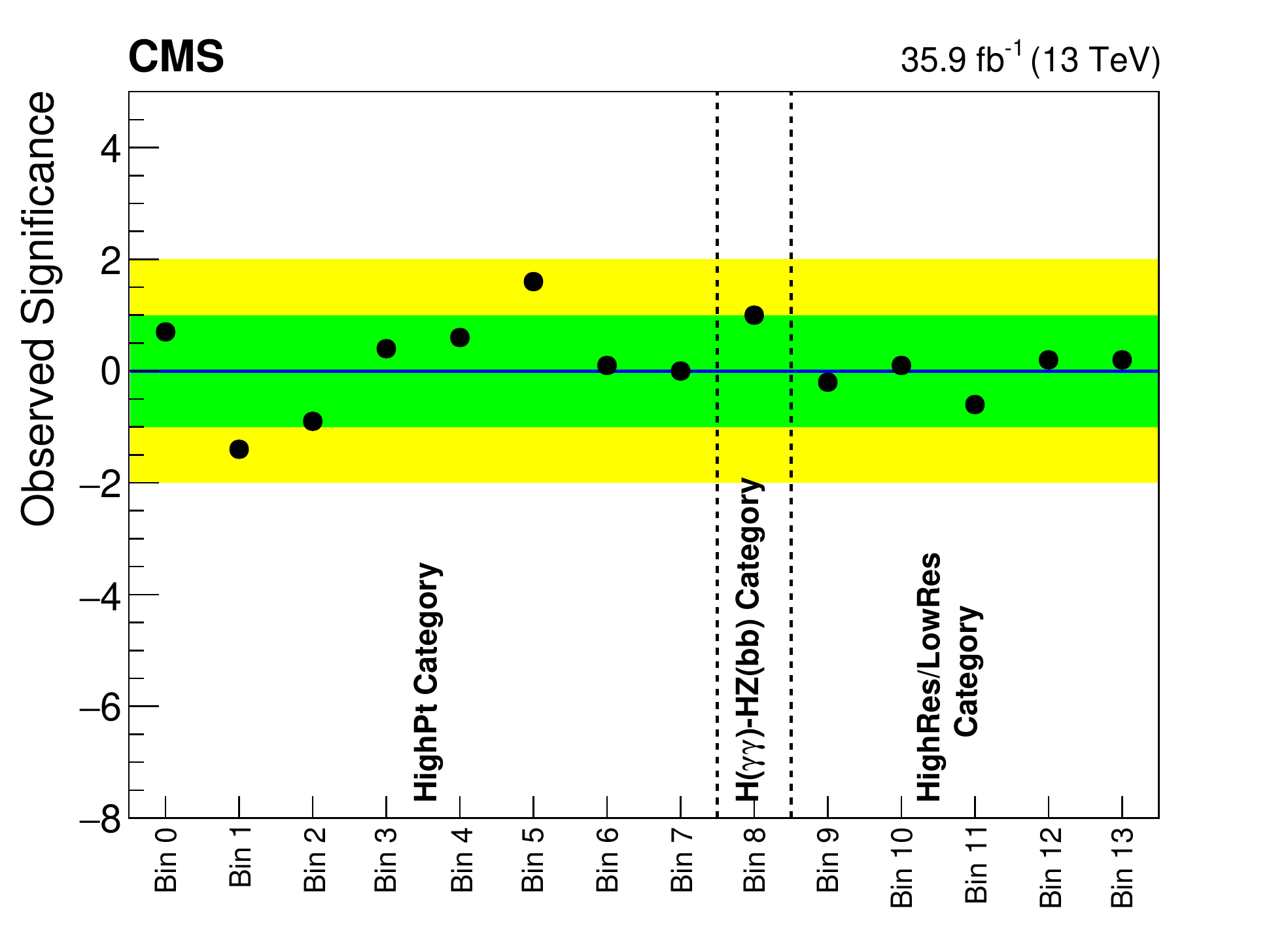}
\caption{
The observed significance in units of standard deviations
is plotted for each search bin. The significance is computed using the profile likelihood, where the sign
reflects whether an excess (positive sign) or deficit (negative sign) is observed.
The categories that the bins belong to are labeled at the bottom. The bins in the HighRes and LowRes categories
are fitted simultaneously and yield a single combined significance.
The yellow and green bands represent the $\pm$1 and $\pm$2 standard deviation regions, respectively.
\label{fig:Significance}}
\end{figure}

The fit results for all search region bins are summarized in
Table~\ref{tab:bkgYieldTable}, along with the data yields, fitted background, and signal yields. An example fit result for the search bin with $\MR > 600$\GeV and $\Rtwo > 0.025$ in the HighPt category is shown in Fig.~\ref{fig:ExampleFit}.
The observed signal significance in each bin is summarized in Fig.~\ref{fig:Significance} for all the search region bins, which are statistically independent. None of the 14 bins exhibits a deviation from the background expectation larger than two standard deviations.

\begin{table*}[htb]
\centering
\topcaption{The nonresonant background yields,
SM Higgs boson background yields, best fit signal yields,
and observed local significance in units of standard deviations ($\sigma$) are shown for the
signal plus background fit in each search region bin. The SM Higgs boson background
yields are slightly altered with respect to the pre-fit predictions from Table~\ref{tab:SMHBkgPrediction}.
The uncertainties include both
statistical and systematic components. The nonresonant background yields correspond to
the yield within the mass window between 122 and 129\GeV
and are intended to estimate the background under the signal peak. The observed
significance for the bins in HighRes and LowRes categories are identical because they
are the result of a simultaneous fit. The significance is computed using the profile
likelihood, where the sign reflects whether an excess (positive sign) or deficit
(negative sign) is observed.
}
\newcolumntype{x}{D{,}{\,\pm\,}{4.2}}
\begin{tabular}{ccxxxc}
\hline
           &          &                   & \multicolumn{1}{c}{Yields}  &                   & Obs. local    \\ \cline{3-5}
Bin        & Category &  \multicolumn{1}{c}{Nonresonant bkg} &  \multicolumn{1}{c}{SM Higgs} &   \multicolumn{1}{c}{Best fit signal}  & significance ($\sigma$)   \\
\hline
0 & HighPt & 36,2      & 6.1,1.0 & 4.8,6.7  & 0.7  \\\
1 & HighPt & 37,2      & 4.3,0.7 & -11,6    & -1.4                  \\
2 & HighPt & 24,2      & 8.4,1.6 & -5.1,5.3 & -0.9                  \\
3 & HighPt & 790,27    & 74,21   & 14,41    & 0.4                   \\
4 & HighPt & 160,15    & 24,6    & 10,15    & 0.6                   \\
5 & HighPt & 34,2      & 5.2,1.2 & 12,8     & 1.6                   \\
6 & HighPt & 127,3     & 29,5    & 1.1,7.9  & 0.1                   \\
7 & HighPt & 40,3      & 7.4,2.0 & -0.3,7.4 & -0.0
\\[\cmsTabSkip]
8 & $\PH(\gamma\gamma)$-$\PH\Z(\PQb\PQb)$ & 65,3 & 2.6,0.4 & 8,8 & 1.0
\\[\cmsTabSkip]
\multirow{2}{*}{9}  & HighRes  & 1792,17     & 77,24   & -9,44 & \multirow{2}{*}{-0.2} \\
                    & LowRes   & 2108,28     & 43,17   & -4,19 &                      \\
\multirow{2}{*}{10} & HighRes  & 44,3        & 1.9,0.6 & 1,8   & \multirow{2}{*}{0.1} \\
                    & LowRes   & 68,3        & 1.0,0.3 & 0,3   &                      \\
\multirow{2}{*}{11} & HighRes  & 127,4       & 5.2,1.4 & -8,12 & \multirow{2}{*}{-0.6}\\
                    & LowRes   & 158,10      & 3.0,1.2 & -4,7  &                      \\
\multirow{2}{*}{12} & HighRes  & 1066,19     & 51,16   & 10,34 & \multirow{2}{*}{0.2} \\
                    & LowRes   & 1310,14     & 29,9    & 5,18  &                      \\
\multirow{2}{*}{13} & HighRes  & 151,5       & 9.5,3.1 & 2,11  & \multirow{2}{*}{0.2} \\
                    & LowRes   & 193,5       & 5.8,2.1 & 1,6   &                     \\
\hline
\end{tabular}
\label{tab:bkgYieldTable}
\end{table*}

We interpret the search results in terms of limits on the production cross section
times branching ratio for simplified models of bottom squark pair-production and chargino-neutralino
production. Diagrams of these simplified models are shown in Fig.~\ref{fig:SMSDiagrams}.
In the case of bottom squark pair production, we consider the scenario where the bottom squark
decays to a bottom quark and the next-to-lightest neutralino ($\PSGczDt$), and the $\PSGczDt$
decays to a Higgs boson and the LSP ($\PSGczDo$), and the production cross sections are computed at
NLO plus next-to-leading-log (NLL) precision with all the other sparticles assumed to be heavy and
decoupled~\cite{NLONLL1,NLONLL2,NLONLL3,NLONLL4,NLONLL5,Borschensky:2014cia}.
We restrict ourselves to the scenario where
the mass splitting between the $\PSGczDt$ and the $\PSGczDo$ is 130\GeV, slightly above threshold
to produce an on-shell Higgs boson. In the case of chargino-neutralino
production, we consider two different scenarios. In the first one, pure wino-like
charginos ($\PSGcpmDo$) and the next-to-lightest neutralino $\PSGczDt$
are mass-degenerate and are produced together, with the chargino decaying to a W boson and the
LSP ($\PSGczDo$) and the $\PSGczDt$ decaying to a Higgs boson and the
LSP ($\PSGczDo$). The production cross sections are computed at NLO plus next-to-leading-log (NLL) precision in a limit
of mass-degenerate wino $\chiz_2$ and $\chipm_1$, light bino $\chiz_1$, and with all the other sparticles
assumed to be heavy and decoupled~\cite{Beenakker:1999xh,Fuks:2012qx,Fuks:2013vua}.
In the second scenario, we consider a GMSB~\cite{Dimopoulos:1996vz,Matchev:1999ft}
simplified model where Higgsino-like charginos and neutralinos are nearly mass-degenerate and are produced in pairs
through the following combinations: $\PSGczDo\PSGczDt$,
$\PSGczDo\PSGcpmDo$, $\PSGczDt\PSGcpmDo$, and $\PSGcpmDo\PSGcmpDo$. Because of the mass degeneracy,
both the $\PSGczDt$ and the $\PSGcpmDo$ will decay to $\PSGczDo$ and other low-$\pt$ (soft) particles, leading to a signature with a
$\PSGczDo$ pair. Each $\PSGczDo$ will subsequently decay to a Higgs boson and the goldstino ($\PXXSG$),
which is the LSP, or to a Z boson and the goldstino. We consider the case where the branching
fraction of the $\PSGczDo\to \PH\PXXSG$ decay is 100\%, and the case where the branching fraction
of the $\PSGczDo\to \PH\PXXSG$ and $\PSGczDo\to \Z\PXXSG$ decays are
each 50\%. The cross sections for higgsino pair production are computed at NLO plus
NLL precision in a limit of mass-degenerate higgsino $\PSGczDt$,
$\PSGcpmDo$, and $\PSGczDo$ with all the other sparticles assumed to be heavy
and decoupled~\cite{Beenakker:1999xh,Fuks:2012qx,Fuks:2013vua}. Following the convention of real mixing matrices and signed
neutralino or chargino masses~\cite{Skands:2003cj}, we set the mass of $\PSGczDo$ ($\PSGczDt$)
to positive (negative) values. The product of the third and fourth
elements of the corresponding rows of the neutralino mixing matrix $N$ is $+0.5$ ($-0.5$).
The elements $U_{12}$ and $V_{12}$ of the chargino mixing matrices are set to 1.

\begin{figure}[htb]
\centering
\includegraphics[width=0.48\textwidth,angle=0.]{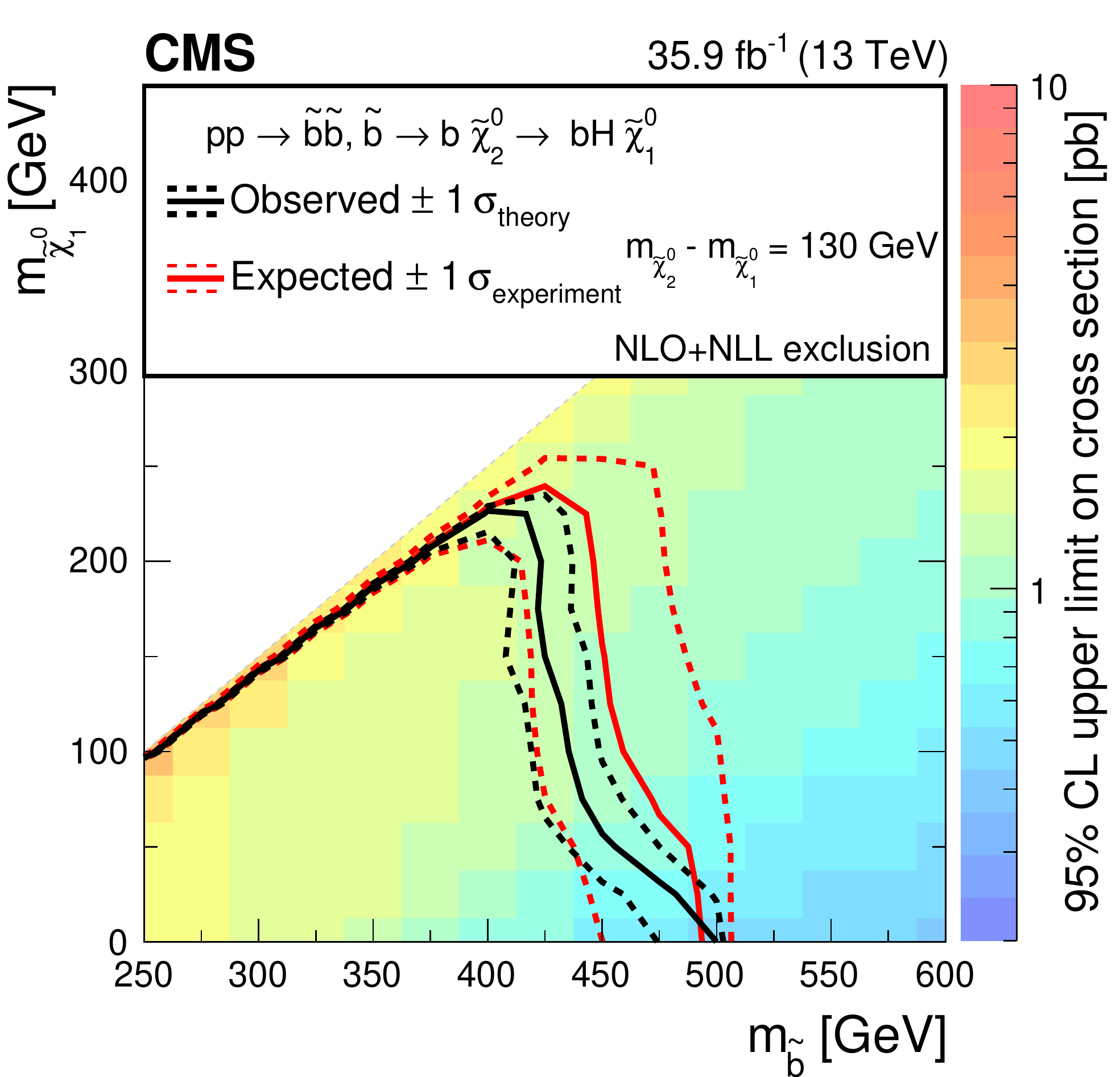}
\includegraphics[width=0.48\textwidth,angle=0.]{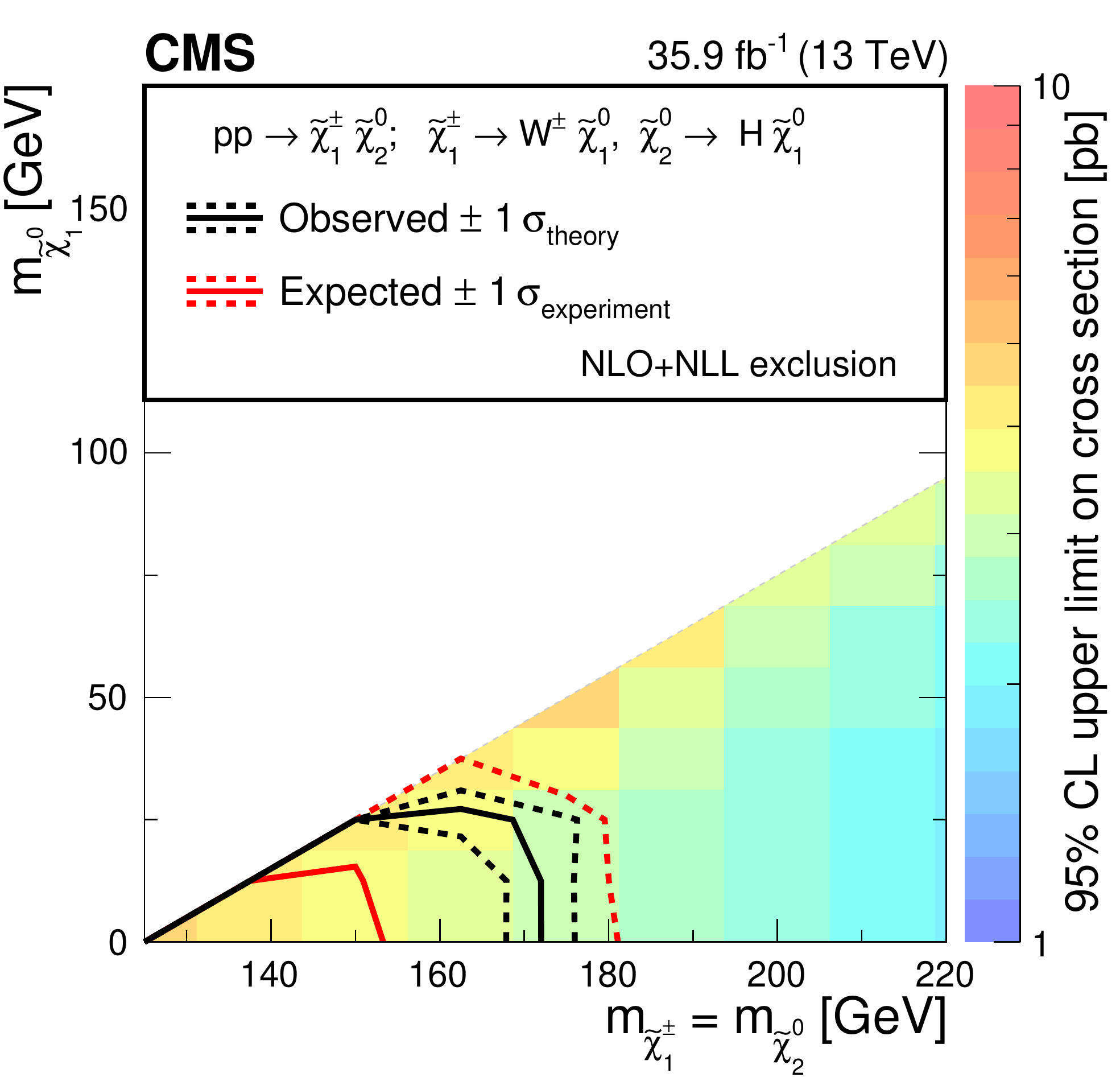}
\caption{
The observed 95\% CL upper limits on the bottom squark pair production cross section (\cmsLeft)
and wino-like chargino-neutralino production cross section (\cmsRight) are shown.
The solid and dotted black contours represent the observed exclusion region and its $\pm$1 standard deviations ($1\sigma$)
of their experimental and theoretical uncertainties,
while the analogous red contours represent the expected exclusion region and its $1\sigma$ band.
\label{fig:LimitsSbottom}}
\end{figure}

\begin{figure}
\centering
\includegraphics[width=0.48\textwidth,angle=0.]{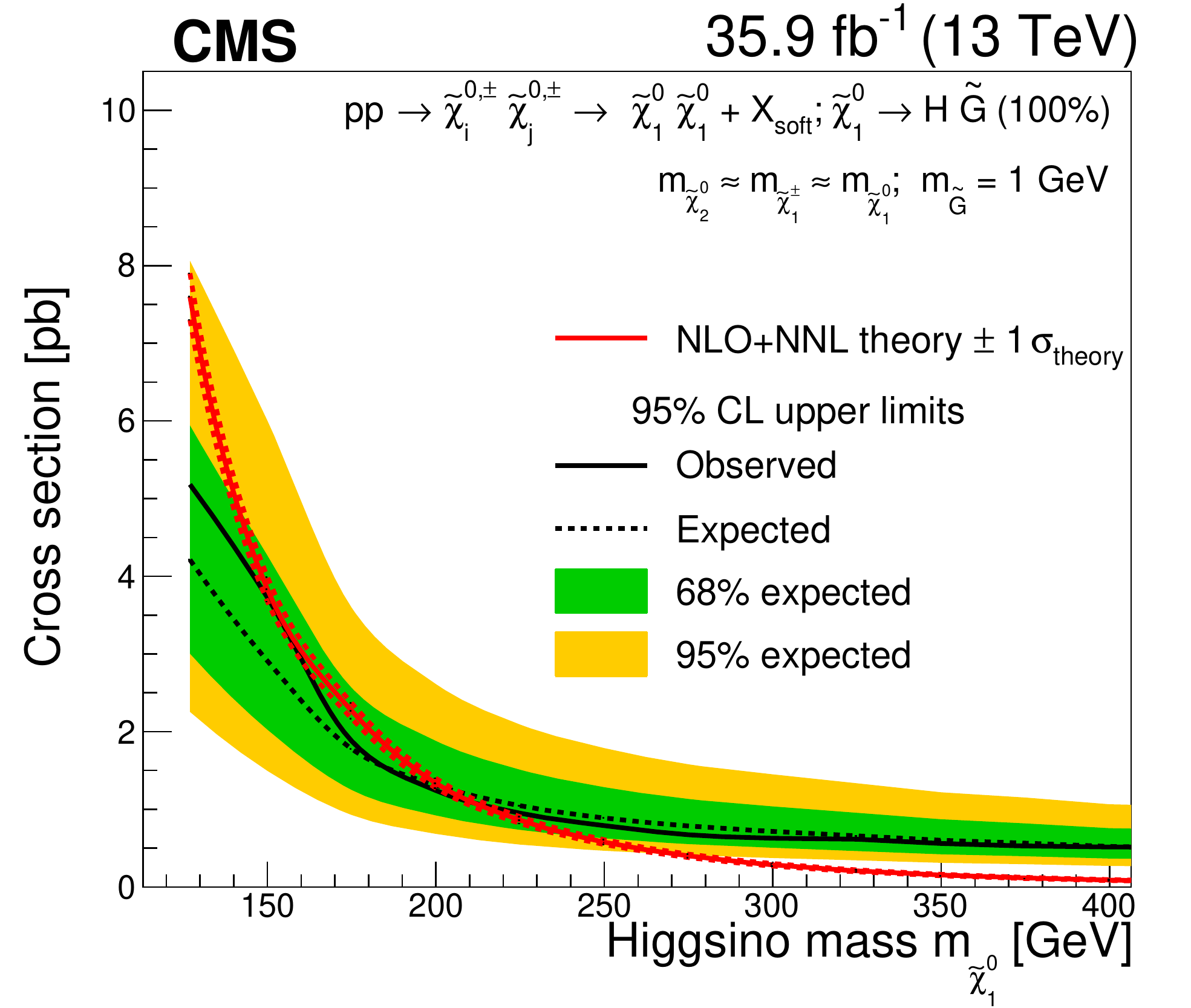}
\includegraphics[width=0.48\textwidth,angle=0.]{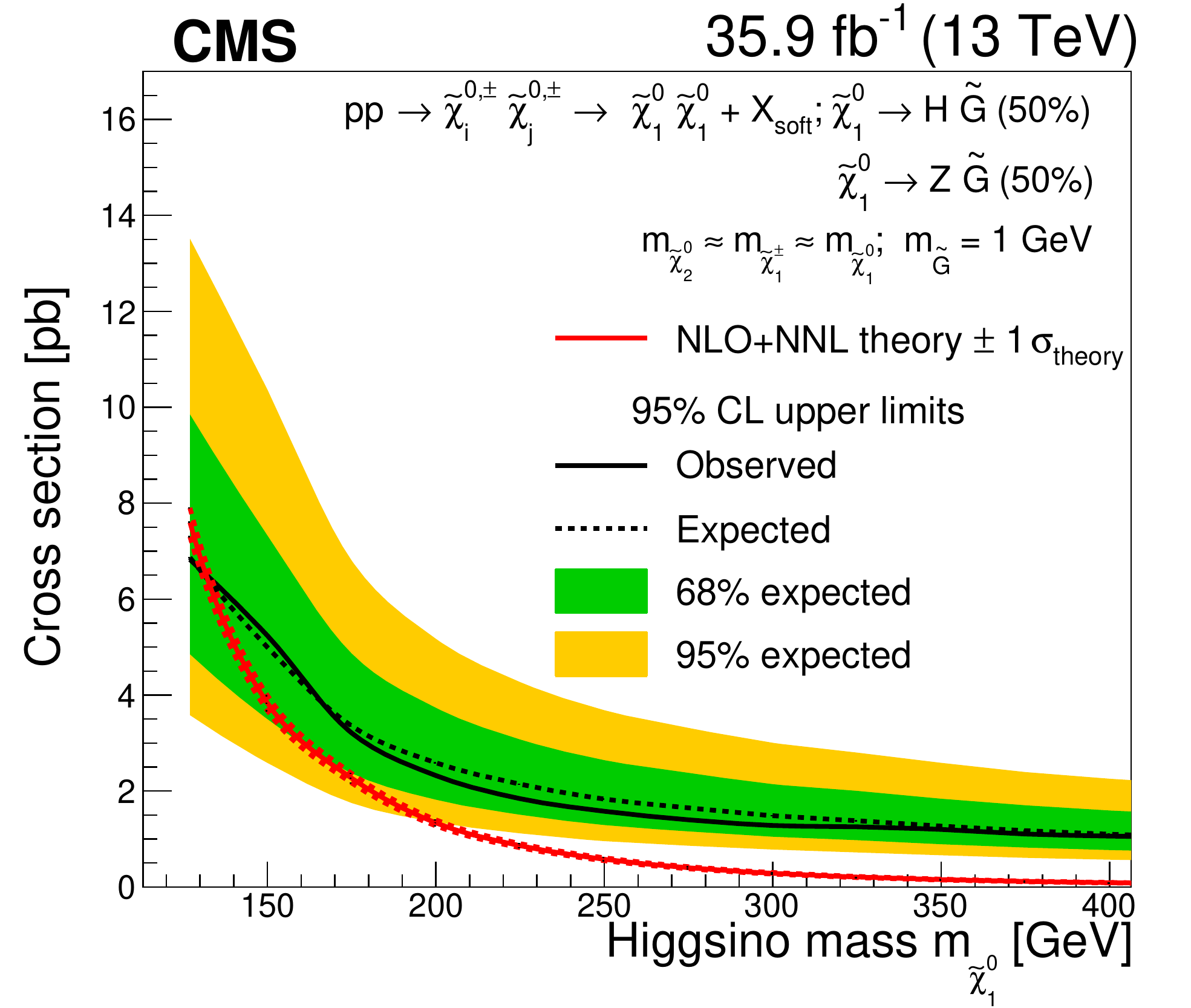}
\caption{ The observed 95\% CL upper limits on the production cross section for higgsino-like chargino-neutralino production are shown. The charginos and neutralinos undergo several cascade decays producing either Higgs or Z bosons. We present limits in the scenario where the branching fraction of the $\PSGczDo\to \PH\PXXSG$ decay is 100\% (\cmsLeft) and the scenario where the branching fraction of the $\PSGczDo\to \PH\PXXSG$ and $\PSGczDo\to \Z\PXXSG$ decays are each 50\% (\cmsRight). The dotted and solid black curves represent the expected and observed exclusion region, and the green and yellow bands represent the $\pm$1 and $\pm$2 standard deviation regions, respectively. The red solid and dotted lines show the theoretical production cross section and its uncertainty band. \label{fig:LimitsTChiHH}}
\end{figure}

Following the $\mathrm{CL}_{\mathrm{s}}$ procedure~\cite{Read:2002hq,Junk:1999kv,ATLAS:2011tau}, we use
the profile likelihood ratio test statistic and the asymptotic
formula~\cite{Cowan:2010js} to evaluate the 95\%~confidence level (CL) observed and expected limits on the signal production cross sections.  For the bottom squark pair production model, the limits are shown on the left of Fig.~\ref{fig:LimitsSbottom} as a function of the bottom squark mass and the LSP mass. We exclude bottom squarks with masses below about 450\GeV for all LSP masses below 250\GeV. For the wino-like chargino-neutralino production
simplified model, the limits are shown on the right of Fig.~\ref{fig:LimitsSbottom} as a function of the chargino mass and the LSP mass. We exclude chargino masses below about
170\GeV for all LSP masses below 25\GeV. For the higgsino-like chargino-neutralino production simplified models, the limits are shown in Fig.~\ref{fig:LimitsTChiHH} as a function
of the chargino mass for the case where the branching fraction of the $\PSGczDo\to \PH\PXXSG$ decay is 100\% on the left, and
for the case where the branching fraction of the $\PSGczDo\to \PH\PXXSG$
and $\PSGczDo\to \Z\PXXSG$ decays are both 50\%, on the right.
We exclude charginos below 205\GeV and 130\GeV in the former and latter cases, respectively.

\section{Summary}\label{sec:summary}

A search for anomalous Higgs boson production through decays of supersymmetric particles is performed with the
proton-proton collision data collected in 2016 by the CMS experiment
at the LHC. The sample corresponds to an integrated
luminosity of 35.9\fbinv at the center-of-mass energy $\sqrt{s}=13$\TeV.
Higgs boson candidates are reconstructed from pairs of photons in the central part of the detector.
The razor variables \MR and \Rtwo are used to suppress Standard Model (SM) Higgs boson production and other SM backgrounds.
The non-resonant background is estimated through a fit to the diphoton mass distribution in data, while
the SM Higgs background is predicted using simulation.
We interpret the results in terms of production cross section limits on simplified models of
bottom squark pair production and chargino-neutralino production. We exclude bottom squark masses below 450\GeV for bottom squarks
decaying to a bottom quark, a Higgs boson, and the lightest supersymmetric particle (LSP) for LSP masses below 250\GeV and assuming
a mass splitting between the $\PSGczDt$ and $\PSGczDo$ of $130$~GeV.
For wino-like chargino-neutralino production, we improved the search sensitivity by a factor of two with respect to
previous results~\cite{Khachatryan:2014mma} and we exclude charginos with mass
below 170\GeV for LSP masses below 25\GeV. In the GMSB scenario, we exclude charginos with
mass below 205\GeV for neutralinos decaying to a Higgs boson and
a goldstino LSP~($\PXXSG$) with 100\% branching fraction. Finally, we exclude charginos with mass below 130\GeV for the
case where the branching fractions of the $\PSGczDo\to \PH\PXXSG$ and
$\PSGczDo\to \Z\PXXSG$ decays are 50\% each.

\begin{acknowledgments}
We congratulate our colleagues in the CERN accelerator departments for the excellent performance of the LHC and thank the technical and administrative staffs at CERN and at other CMS institutes for their contributions to the success of the CMS effort. In addition, we gratefully acknowledge the computing centers and personnel of the Worldwide LHC Computing Grid for delivering so effectively the computing infrastructure essential to our analyses. Finally, we acknowledge the enduring support for the construction and operation of the LHC and the CMS detector provided by the following funding agencies: BMWFW and FWF (Austria); FNRS and FWO (Belgium); CNPq, CAPES, FAPERJ, and FAPESP (Brazil); MES (Bulgaria); CERN; CAS, MoST, and NSFC (China); COLCIENCIAS (Colombia); MSES and CSF (Croatia); RPF (Cyprus); SENESCYT (Ecuador); MoER, ERC IUT, and ERDF (Estonia); Academy of Finland, MEC, and HIP (Finland); CEA and CNRS/IN2P3 (France); BMBF, DFG, and HGF (Germany); GSRT (Greece); OTKA and NIH (Hungary); DAE and DST (India); IPM (Iran); SFI (Ireland); INFN (Italy); MSIP and NRF (Republic of Korea); LAS (Lithuania); MOE and UM (Malaysia); BUAP, CINVESTAV, CONACYT, LNS, SEP, and UASLP-FAI (Mexico); MBIE (New Zealand); PAEC (Pakistan); MSHE and NSC (Poland); FCT (Portugal); JINR (Dubna); MON, RosAtom, RAS, RFBR and RAEP (Russia); MESTD (Serbia); SEIDI, CPAN, PCTI and FEDER (Spain); Swiss Funding Agencies (Switzerland); MST (Taipei); ThEPCenter, IPST, STAR, and NSTDA (Thailand); TUBITAK and TAEK (Turkey); NASU and SFFR (Ukraine); STFC (United Kingdom); DOE and NSF (USA).

\hyphenation{Rachada-pisek} Individuals have received support from the Marie-Curie program and the European Research Council and Horizon 2020 Grant, contract No. 675440 (European Union); the Leventis Foundation; the A. P. Sloan Foundation; the Alexander von Humboldt Foundation; the Belgian Federal Science Policy Office; the Fonds pour la Formation \`a la Recherche dans l'Industrie et dans l'Agriculture (FRIA-Belgium); the Agentschap voor Innovatie door Wetenschap en Technologie (IWT-Belgium); the Ministry of Education, Youth and Sports (MEYS) of the Czech Republic; the Council of Science and Industrial Research, India; the HOMING PLUS program of the Foundation for Polish Science, cofinanced from European Union, Regional Development Fund, the Mobility Plus program of the Ministry of Science and Higher Education, the National Science Center (Poland), contracts Harmonia 2014/14/M/ST2/00428, Opus 2014/13/B/ST2/02543, 2014/15/B/ST2/03998, and 2015/19/B/ST2/02861, Sonata-bis 2012/07/E/ST2/01406; the National Priorities Research Program by Qatar National Research Fund; the Programa Clar\'in-COFUND del Principado de Asturias; the Thalis and Aristeia programs cofinanced by EU-ESF and the Greek NSRF; the Rachadapisek Sompot Fund for Postdoctoral Fellowship, Chulalongkorn University and the Chulalongkorn Academic into Its 2nd Century Project Advancement Project (Thailand); the Welch Foundation, contract C-1845; and the Weston Havens Foundation (USA).
\end{acknowledgments}

\clearpage
\bibliography{auto_generated}

\cleardoublepage \appendix\section{The CMS Collaboration \label{app:collab}}\begin{sloppypar}\hyphenpenalty=5000\widowpenalty=500\clubpenalty=5000\textbf{Yerevan Physics Institute,  Yerevan,  Armenia}\\*[0pt]
A.M.~Sirunyan, A.~Tumasyan
\vskip\cmsinstskip
\textbf{Institut f\"{u}r Hochenergiephysik,  Wien,  Austria}\\*[0pt]
W.~Adam, F.~Ambrogi, E.~Asilar, T.~Bergauer, J.~Brandstetter, E.~Brondolin, M.~Dragicevic, J.~Er\"{o}, M.~Flechl, M.~Friedl, R.~Fr\"{u}hwirth\cmsAuthorMark{1}, V.M.~Ghete, J.~Grossmann, J.~Hrubec, M.~Jeitler\cmsAuthorMark{1}, A.~K\"{o}nig, N.~Krammer, I.~Kr\"{a}tschmer, D.~Liko, T.~Madlener, I.~Mikulec, E.~Pree, D.~Rabady, N.~Rad, H.~Rohringer, J.~Schieck\cmsAuthorMark{1}, R.~Sch\"{o}fbeck, M.~Spanring, D.~Spitzbart, J.~Strauss, W.~Waltenberger, J.~Wittmann, C.-E.~Wulz\cmsAuthorMark{1}, M.~Zarucki
% ...
\vskip\cmsinstskip
% ...
1:~~Also at Vienna University of Technology, Vienna, Austria\\
% ...

\end{sloppypar}
\end{document}